\documentclass[pra,amsmath,amssymb,showpacs,twocolumn]{revtex4}
 \usepackage{graphicx}
 \usepackage[latin1]{inputenc}
 \usepackage{xspace}
 \usepackage{latexsym}
 \usepackage{amsmath}
 \usepackage{amssymb}
 \usepackage{amstext}
 \usepackage{amsxtra}
 \usepackage{makeidx}
 
 \newcommand{\bs}{\boldsymbol}

 \numberwithin{equation}{section}

\def\O{\Omega}

\def\fint{\int\!\!\!\!{\hbox{\tiny$^\diagup$}}}
\def\ffint{\int\!\!\!\!\!{\hbox{\tiny$^\diagup$}}}

\def\r0{\rho_{0}}

\def\a0{\alpha_0}
\def\a{\alpha}

\def\g{\gamma}
\def\s{\sigma}

\def\la{\lambda}

\def\be{\begin{equation}}
\def\ee{\end{equation}}
\def\beq{\begin{equation}}
\def\eeq{\end{equation}}

\def\O{\Omega}

\def\({\left(}
\def\){\right)}

\begin{document}

\title{Vortex patterns in a fast rotating Bose-Einstein condensate}
\author{Amandine Aftalion, Xavier Blanc}
\affiliation{Laboratoire Jacques-Louis Lions, Universit{\'e} Paris 6,\\ 175 rue du Chevaleret, 75013 Paris, France. }
\author{Jean Dalibard}
\affiliation{Laboratoire Kastler Brossel, 24 rue Lhomond, 75005
Paris, France}

\date{\today}


 \begin{abstract}
For a fast rotating condensate in a harmonic trap, we investigate
the structure of the vortex lattice using wave functions
minimizing the Gross Pitaveskii energy in the Lowest Landau Level.
We find that the minimizer of the energy in the rotating frame has
a distorted vortex lattice for which we plot the typical
distribution. We  compute analytically the energy of an infinite
regular lattice and of a class of distorted lattices. We find the
optimal distortion and relate it to the decay of the wave
function. Finally, we  generalize our method to other trapping
potentials.
 \end{abstract}


\maketitle


The rotation of a macroscopic quantum fluid is a source of
fascinating problems. By contrast with a classical fluid, for
which the equilibrium velocity field corresponds to rigid body
rotation, a quantum fluid described by a macroscopic wave function
rotates through the nucleation of quantized vortices
\cite{Lifshitz,Donnelly91}. A vortex is a singular point (in 2
dimensions) or line (in 3 dimensions) where the density vanishes.
Along a contour encircling a vortex, the circulation of the
velocity is quantized in units of $h/m$, where $m$ is the mass of
a particle of the fluid.

Vortices are universal features which appear in many macroscopic
quantum systems, such as superconductors or superfluid liquid
helium. Recently, detailed investigations have been performed on
rotating atomic gaseous Bose-Einstein condensates. These
condensates are usually confined in a harmonic potential, with
cylindrical symmetry around the rotation axis $z$. Two limiting
regimes occur depending on the ratio of the rotation frequency
$\Omega$ and the trap frequency $\omega$ in the $xy$ plane. When
$\Omega$ is notably smaller than $\omega$, only one or a few
vortices are present at equilibrium \cite{Madison00,Matthews99}.
When $\Omega$ approaches $\omega$, since the centrifugal force
nearly balances the trapping force, the radius of the rotating gas
increases and tends to infinity, and the number of vortices in the
condensate diverges
\cite{Ketterle1,Boulder03,Boulder04,Boulder04bis}.

As pointed out by several authors, the fast rotation regime
presents a strong analogy with Quantum Hall physics. Indeed the
one-body hamiltonian written in the rotating frame is similar to
that of a charged particle in a uniform magnetic field. Therefore
the ground energy level is macroscopically degenerate, as the
celebrated Landau levels obtained for the quantum motion of a
charge in a magnetic field. There are two aspects in this
connection with Quantum Hall Physics. Firstly, when the number of
vortices inside the fluid remains small compared to the number $N$
of atoms, we expect that the ground state of the system will
correspond to a Bose-Einstein condensate, described by a
macroscopic wave function $\psi(\bs r)$. This situation has been
referred to as `mean field Quantum Hall regime'
\cite{H,Fischer03,BP,WP,WBP,KCR}. Secondly, when $\Omega$
approaches $\omega$ even closer, the number of vortices reaches
values comparable to the total number of atoms $N$. The
description by a single macroscopic wave function then breaks
down, and one expects a strongly correlated ground state, such as
that of an electron gas in the fractional quantum Hall regime
\cite{Cooper01,Paredes01,Sinova02,Reijnders02,Regnault03}. We do
not address the second situation in this paper and we rather focus
on the first regime. Furthermore we restrict our analysis to the
case of a two-dimensional gas in the $xy$ plane, assuming that a
strong confinement along the $z$ direction so that the
corresponding degree of freedom is frozen.

The main features of the vortex assembly equilibrium in the fast
rotation regime are well known. The vortices form a triangular
Abrikosov lattice in the $xy$ plane and the area of the elementary
cell is ${\cal A}=\pi \hbar/(m\Omega)$ \cite{Feynman}. The atomic
velocity field obtained by a coarse-grained average over a few
elementary cells is equal to the rigid body rotation result $\bs
v=\bs \Omega \times \bs r$, where $\bs \Omega=\Omega \hat {\bs z}$
($\hat {\bs z}$ is the unit vector along the $z$ axis).

Beyond this approximation, the physics is very rich and many
points are still debated. In a seminal paper \cite{H}, Ho
introduced the description of the macroscopic state of the
rotating gas in the $xy$ plane by a wave function belonging to the
Lowest Landau Level (LLL). An LLL wave function is entirely
determined (up to a global phase factor) by the location of
vortices. Ho considered the case of a uniform infinite vortex
lattice and inferred that the ground state of the system
corresponds to a gaussian shape for the coarse-grained atom
density profile. Using both analytical \cite{BP,WP,WBP} and
numerical \cite{KCR} investigations, it was subsequently pointed
out that the atom distribution may have the shape of an inverted
parabola, instead of Ho's gaussian result. Two paths have been
proposed to explain the emergence of such non gaussian profiles.
The first one assumes that the restriction to the LLL is not
sufficient and the contamination of the ground state wave function
by other Landau levels is responsible for the transition from a
gaussian to an inverted parabola \cite{BP,WP}. The second path
explores the influence of distortions of the vortex lattice
(within the LLL) to account for the deviation of the equilibrium
profile from a gaussian \cite{WBP,KCR}.

\begin{figure}[bt]
\begin{centering}
\includegraphics[width=85mm]{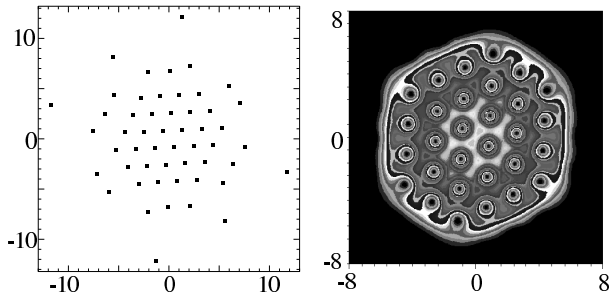}
\end{centering}
 \caption{Structure of the ground state of a rotating Bose-Einstein
 condensate described by a LLL wavefunction ($\Lambda=3000)$.
(a) Vortex location; (b) Atomic density profile (with a larger
scale); The reduced energy defined in Eq.~(\ref{energy3}) is
$\epsilon=31.4107101$.}
 \label{fig:Photovx}
\end{figure}

In the present paper, we investigate the structure of the vortex
lattice for a fast rotating condensate. We derive the condition
under which the LLL is a proper variational space to determine the
ground state wave function within a good approximation. We present
a numerical and analytical analysis of the structure of the vortex
lattice, based on a minimization of the Gross-Pitaevskii energy
functional within the LLL. We find that the vortices lie in a
bounded domain, and that the lattice is strongly distorted on the
edges of the domain.  This leads to a breakdown of the rigid body
rotation hypothesis which, as said above, would correspond to a
uniform infinite lattice with a prescribed volume of the cell. The
distortion of the vortex lattice is such that, in a harmonic
potential, the coarse-grained average of the atomic density varies
as an inverted parabola over the region where it takes significant
values (Thomas-Fermi distribution). A similar conclusion has also
been reached recently in \cite{WBP,KCR}.  In addition to the
atomic density profile, our numerical computations give access to
the exact location of the zeroes of the wave function, i.e. the
vortices.

An example of relevant vortex and atom distributions is shown in
Fig.~\ref{fig:Photovx}a and Fig.~\ref{fig:Photovx}b for $n=52$
vortices. The parameters used to obtain this vortex structure
correspond to a quasi-two dimensional gas of  $1000$ rubidium
atoms, rotating in the $xy$ plane at a frequency
$\Omega=0.99\,\omega$, and strongly confined along the $z$ axis
with a trapping frequency $\omega_z/(2\pi)=150$~Hz. The spatial
distribution of vortices corresponds to the triangular Abrikosov
lattice only around the center of the condensate: there are about
30 vortices on the quasi-regular part of the lattice and they lie
in the region where the atomic density is significant: these are
the only ones seen in the density profile of
Fig.~\ref{fig:Photovx}b. At the edge of the condensate, the atomic
density is  reduced with respect to the central density, the
vortex surface density drops down, and the vortex lattice is
strongly distorted. Our analytical approach allows to justify this
distortion and its relationship with the decay of the solution.

The paper is organized as follows. We start
(\S~\ref{sec:singlepart}) with a short review of the energy levels
of a single, harmonically trapped particle in a rotating frame,
and we give the expression of the Landau levels for the problem of
interest. Then, we  consider the problem of an interacting gas in
rotation, and we derive the condition for this gas to be well
described by an LLL wave function (\S~\ref{sec:interacting}).
Sections \ref{sec:eqLLL} and \ref{sec:extension} contain the main
original results of the paper. In \S~\ref{sec:eqLLL}, we explain
how to improve the determination of the ground state energy by
relaxing the hypothesis of an infinite regular lattice. We present
analytical estimates for an LLL wave function with a distorted
vortex lattice, and we show that these estimates are in excellent
agreement with the results of the numerical approach. In
\S~\ref{sec:extension} we  extend the method to non harmonic
confinement, with the example of a quadratic+quartic potential.
Finally we give in section \ref{sec:conclusion} some conclusions
and perspectives.


\section{Single particle physics in a rotating frame}
\label{sec:singlepart}

In this section, we briefly review the main results concerning the
energy levels of a single particle confined in a two-dimensional
isotropic harmonic potential of frequency $\omega$ in the $xy$
plane. We are interested here in the energy level structure in the
frame rotating at angular frequency $\Omega$ ($>0$) around the $z$
axis, perpendicular to the $xy$ plane.

In the following, we choose $\omega$, $\hbar \omega$, and
$\sqrt{\hbar/(m\omega)}$, as units of frequency, energy and
length, respectively. The hamiltonian of the particle is
 \begin{eqnarray}
H^{(1)}_\Omega &=&-\frac{1}{2} \nabla^2 + \frac{r^2}{2} -\Omega
L_z \nonumber
\\
&=&-\frac{1}{2}\left(\bs \nabla -i\bs A \right)^2 + (1-\Omega^2)
\frac{r^2}{2}
 \label{singlepartH}
 \end{eqnarray}
with $r^2=x^2+y^2$ and $\bs A=\bs \Omega \times \bs r$.  This
energy is the sum of three terms: kinetic energy, potential energy
$r^2/2$, and `rotation energy' $-\Omega L_z$ corresponding to the
passage in the rotating frame. The operator
$L_z=i(y\partial_x-x\partial_y)$ is the $z$ component of the
angular momentum.

\subsection{The Landau level structure}

\begin{figure}
 \centerline{\includegraphics[width=7cm]{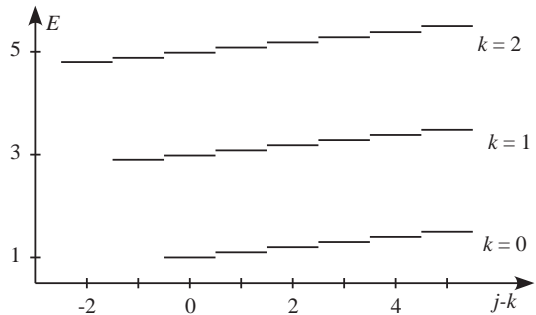}}
 \caption{Single particle spectrum for $\Omega=0.9$. The index $k$
labels the Landau levels. }
 \label{fig:LL}
\end{figure}

Eq. (\ref{singlepartH}) is formally identical to the hamiltonian
of a particle of charge 1 placed in a uniform magnetic field
$2\Omega \bs {\hat z}$, and confined in a potential with a spring
constant $1-\Omega^2$. A common eigenbasis of $L_z$ and $H$ is the
set of (not normalized) Hermite functions:
  \begin{equation}
\phi_{j,k}(\bs r)=e^{r^2/2}\;
(\partial_x+i\partial_y)^j\;(\partial_x-i\partial_y)^k
\left(e^{-r^2}\right)
 \label{Hermite}
  \end{equation}
where $j$ and $k$ are non-negative integers. The eigenvalues are
$j-k$ for $L_z$ and
 \begin{equation}
 E_{j,k}=1+ (1-\Omega)j+ (1+\Omega)k
\label{spectrum}
 \end{equation}
for $H$. For $\Omega=1$, these energy levels group in series of
states with a given $k$, corresponding to the well known Landau
levels. Each Landau level has an infinite degeneracy. For $\Omega$
slightly smaller than 1, this structure in terms of Landau levels
labeled by the index $k$ remains relevant, as shown in Fig.
\ref{fig:LL}. The lowest energy states of two adjacent Landau
levels are separated by $\sim 2$, whereas the distance between two
adjacent states in a given Landau level is $1-\Omega \ll 1$.

It is clear from these considerations that the rotation frequency
$\Omega$ must be chosen smaller than the trapping frequency in the
$xy$ plane, i.e. $\omega=1$ with our choice of units. Otherwise
the single particle spectrum Eq.~(\ref{spectrum}) is not bounded
from below. Physically, this corresponds to the requirement that
the centrifugal force $m\Omega^2 r$ must not exceed the restoring
force in the $xy$ plane $-m\omega^2 r$.

\subsection{The lowest Landau level}

When the rotation frequency $\Omega$ is close to 1, the states of
interest at low temperature are essentially those associated with
$k=0$, e.g. the lowest Landau level (LLL)
\cite{Girvin84,Rokhsar99,H}. Any function $\psi(\bs r)$ of the LLL
is a linear combination of the $\phi_{j,0}$'s and it can be cast
in the form:
 \begin{equation}
\psi(\bs r)=e^{-r^2/2}\; P(u)
 \label{formLLL1}
 \end{equation}
where $\bs r=(x,y)$, $u=x+iy$ and $P(u)$ is a polynomial (or an
analytic function) of $u$. When $P(u)$ is a polynomial of degree
$n$, an alternative form of $\psi(\bs r)$ is
 \begin{equation}
\psi(\bs r)= e^{-r^2/2}\;\prod_{j=1}^n (u-u_j)
 \label{formLLL2}
 \end{equation}
where the $u_j$ ($j=1 \ldots n$) are the $n$ complex zeroes of
$P(u)$. Each $u_j$ is the position of a single-charged, positive
vortex, since the phase of $\psi(\bs r)$ changes by $2\pi$ along a
closed contour encircling $u_j$.

In the LLL, there is a one-to-one correspondence between atom and
vortex distributions. This relation can be made explicit by
introducing the atom density $\rho_a(\bs r)=|\psi(\bs r)|^2$:
 \begin{equation}
\ln(\rho_a(\bs r))= -r^2+ 2\sum_j \ln|\bs r-\bs r_j|\ .
 \end{equation}
Introducing the vortex density $\rho_v(\bs r) = \sum_j \delta(\bs
r-\bs r_j)$ we obtain using $\nabla^2 \left[\ln|\bs r-\bs
r_0|\right]= 2\pi\,\delta(\bs r-\bs r_0)$:
 \begin{equation}
\nabla^2 \left[\ln (\rho_a(\bs r)) \right]=-4+4\pi\,  \rho_v(\bs
r)\ . \label{rhovrhoa}
 \end{equation}
This relation was initially derived by Ho in \cite{H} who
interpreted it in terms of the Gauss law for a system of
two-dimension charges located at the points $\bs r_j$.


\section{The interacting gas in rotation}
\label{sec:interacting}

We now consider a gas of $N$ identical bosonic atoms with mass
$m$. The gas is confined in a cylindrically symmetric harmonic
potential, with frequency $\omega$ in the $xy$ plane and
$\omega_z$ along the $z$ direction. We suppose that the
characteristic energy $\hbar \omega_z$ is very large compared to
all other energy scales appearing in the paper, so that we can
assume that the atoms occupy the ground state of the $z$ motion,
of energy $\hbar \omega_z/2$ and extension
$a_z=\sqrt{\hbar/(m\omega_z)}$. We are interested in the ground
state of this quasi-two dimensional gas, when it is rotating at
frequency $\Omega$ close to $\omega$ around the $z$ axis.

\subsection{The Gross-Pitaevskii energy functional}

The state of the gas is described by a macroscopic wave function
$\psi(\bs r)$ normalized to unity, which minimizes the
Gross-Pitaevskii energy functional. We introduce the dimensionless
coefficient $G$ characterizing the strength of atomic
interactions, proportional to the atom scattering length $a_s$:
$G=\sqrt{8\pi}N a_s/a_z$. The average energy per atom, written in
the frame rotating at frequency $\Omega$, is:
 \begin{equation}
E[\psi]=\int \left( \psi^* \left[H^{(1)}_\Omega\psi\right] +
\frac{G}{2} |\psi|^4  \right) \; d^2r
 \label{energy1}
 \end{equation}
where $H^{(1)}_\Omega$ is defined in Eq. (\ref{singlepartH}). The
wave function $\psi(\bs r)$ minimizing $E[\psi]$ satisfies the
Gross-Pitaevskii equation:
 \begin{equation}
H^{(1)}_\Omega \psi(\bs r) +G|\psi(\bs r)|^2 \psi(\bs r)=\mu
\,\psi(\bs r)\ .
 \label{GP1}
 \end{equation}
The chemical potential $\mu$ is determined by imposing that
$\int|\psi|^2=1$. The solution of Eq.~(\ref{GP1}) depends on the
two independent dimensionless parameters $\Omega$ and $G$.

\subsection{The LLL limit}

In the presence of repulsive interactions ($G>0$), the basis of
Eq.~(\ref{Hermite}) is not an eigenbasis of the $N$-body
hamiltonian. However for a given interaction strength $G$ and for
a sufficiently fast rotation ($\Omega$ close to $\omega=1$), the
restriction to the LLL is sufficient to determine with a good
accuracy the ground state of the system and its energy. Indeed
when $\Omega$ approaches $\omega=1$, the centrifugal force
$m\Omega^2 r$ nearly compensates the trapping force $-m\omega^2 r$
and the area occupied by the atoms increases. The effect of
interactions gets smaller so that the total energy per particle
tends to the energy $\hbar \omega$ of the lowest Landau level,
i.e. 1 in our reduced units.

When $\psi$ is chosen in the LLL, the energy functional
Eq.~(\ref{energy1}) can be notably simplified. Indeed the LLL
functions satisfy the equalities:
 \begin{equation}
\langle E_{\rm kin} \rangle=\langle E_{\rm ho} \rangle =
\frac{1}{2}+\frac{1}{2}\int \psi^* \left[L_z\psi\right] \; d^2r
 \label{equalities}
 \end{equation}
where the kinetic and harmonic oscillator energies are:
\begin{equation}
\langle E_{\rm kin} \rangle = \frac{1}{2}\int  |\bs \nabla \psi|^2
\; d^2r \qquad  \langle E_{\rm ho} \rangle=\frac{1}{2}\int r^2\,
|\psi|^2 \; d^2r\ .
\end{equation}
The total energy $E[\psi]=E_{\rm LLL}[\psi]$ is then given by
 \begin{equation} E_{\rm
LLL}[\psi]-\Omega = \int \left( (1-\Omega) r^2 |\psi|^2 +
\frac{G}{2} |\psi|^4  \right) \; d^2r\ .
 \label{energy2}
 \end{equation}
In section~\ref{sec:eqLLL}, we will minimize this energy
functional for functions in the LLL. Here we simply outline some
relevant scaling laws in this regime.

\subsection{Scaling laws and lower bound in the LLL}
\label{subsec:scaling}

The minimization of Eq.~(\ref{energy2}) is equivalent to the
minimization of the reduced energy
 \begin{equation}
\epsilon[\psi]=\frac{E_{\rm LLL}[\psi]-\Omega}{1-\Omega} =\int
\left( r^2 |\psi|^2 + \frac{\Lambda}{2} |\psi|^4  \right) \; d^2r
 \label{energy3}
\end{equation}
with
 \begin{equation}
 \Lambda=\frac{G}{1-\Omega}\ .
 \label{Lambda}
 \end{equation}
Therefore the minimizer $\psi_{\rm LLL}$ depends only on the
parameter $\Lambda$. This is quite different from what happens
when the LLL limit is not reached: for the minimization of
Eq.~(\ref{energy1}), the two parameters $G$ and $\Omega$ are
relevant, and not only their combination $\Lambda$.

It is instructive to consider the minimum of $\epsilon[\psi]$ when
$\psi$ is allowed to explore the whole function space of
normalized functions $\int |\psi|^2 \,d^2r=1$. This minimum is
straightforwardly obtained for $|\psi|^2$ varying as an inverted
parabola in the disk of radius $R_0$:
 \begin{equation}
|\psi_{\rm min}(\bs r)|^2=\displaystyle{\frac{2}{\pi
R_0^2}\left(1-\frac{r^2}{R_0^2} \right)}\ ,\quad R_0= \left(
\frac{2\Lambda}{\pi}\right)^{1/4}
 \label{ThomasFermi}
 \end{equation}
and $\psi(\bs r)=0$ outside. The reduced energy is
 \begin{equation}
\epsilon_{\rm min}=\frac{2\sqrt{2}}{3\sqrt{\pi}}\sqrt{\Lambda}\ .
 \label{LLLbound}
 \end{equation}

The variation of the atomic density as an inverted parabola is
very reminiscent of the Thomas-Fermi distribution for a condensate
at rest in a harmonic potential.  However this analogy should be
taken with care. In the usual Thomas-Fermi approach, one neglects
the kinetic energy term and the equilibrium distribution is found
as a balance between potential and interaction energies. In the
LLL problem considered here, kinetic and potential energies are
equal (see Eq.~(\ref{equalities})), and their sum $\int r^2
|\psi|^2$, which is large compared to 1 when $\Lambda \gg 1$, is
nearly balanced by the rotation term $-\Omega \langle L_z\rangle$.

The function $\psi_{\rm min}$ clearly does not belong to the LLL,
since the only LLL function depending solely on the radial
variable is $\exp(-r^2/2)$. Consequently the reduced energy
Eq.~(\ref{LLLbound}) is strictly lower than the result of the
minimization of $\epsilon[\psi]$ with $\psi$ varying only in the
LLL. In other words, the minimization of Eq.~(\ref{energy2}) that
we perform in the next section, amounts to find the LLL function
which is ``the most similar" to $\psi_{\rm min}$, so that its
reduced energy is the closest to $\epsilon_{\rm min}$. For
$\Lambda \gg 1$, we shall see that $\epsilon \simeq \alpha
\sqrt{\Lambda}$, where $\alpha$ is a coefficient of order unity to
be determined.

\subsection{Validity of the LLL approximation}

Since $\epsilon\simeq \alpha \sqrt{\Lambda}$, the ground state
energy $E$ of the fast rotating gas determined within the LLL
approximation is $\O+\alpha\sqrt{G(1-\Omega)}$. Therefore the
restriction to the LLL is valid as long as the excess energy
$\alpha\sqrt{G(1-\Omega)}$ is small compared to the splitting
$2\hbar\omega=2$ between the LLL and the first excited Landau
level:
 \begin{equation}
\mbox{Restriction to LLL if:}\qquad G (1-\Omega) \ll 1
 \label{LLLfrontier}
 \end{equation}
When this condition is satisfied, the projection of $\psi$ on the
excited Landau levels is negligibly small.

It is interesting to compare the scaling laws derived in the LLL
with the exact relations obtained using the virial theorem. For a
2D gas, this theorem gives for the ground state of the (possibly
rotating) system
 \begin{eqnarray}
\langle E_{\rm ho} \rangle &=& \langle E_{\rm kin} \rangle + \langle
E_{\rm int} \rangle \\
\langle E_{\rm int} \rangle &=&\frac{G}{2}\int |\psi|^4\;d^2r
 \end{eqnarray}
while we expect for LLL wavefunctions for $\Lambda\gg 1$
 \begin{equation}
\langle E_{\rm ho} \rangle = \langle E_{\rm kin} \rangle \sim
\sqrt{\Lambda} \gg \langle E_{\rm int} \rangle \sim
\sqrt{G(1-\Omega)}\ .
 \end{equation}
Therefore, within the LLL validity domain of
Eq.~(\ref{LLLfrontier}), the scaling laws for the predicted LLL
energies agree with the constraints imposed by the virial theorem.


\section{The LLL equilibrium distribution}
\label{sec:eqLLL}

This section is devoted to the minimization of the reduced energy
given in Eq.~(\ref{energy3}) for LLL wave functions. We start with
wave functions corresponding to an infinite regular vortex lattice
and we derive the corresponding energy. Then, we  give numerical
results which we use in the rest of the section as a guide to
improve our choice for trial wave functions and analyze the
distortion of the lattice.

\subsection{The case of a regular vortex lattice}
\label{subsec:regular}

\subsubsection{The average density profile for a regular vortex lattice}

We consider a wave function in the LLL with an infinite number of
vortices on a regular lattice and an average spatial density
$\bar \rho_v$. We denote by $u_j$ the points of the regular
triangular lattice, and by ${\cal A}=1/\bar \rho_v$ the area of
its unit cell. We consider the LLL wave functions
 \begin{equation}
\psi(\bs r)=C e^{-r^2/2} \prod_{u_j\in D_R} (u-u_j)\ ,
 \label{truncated}
 \end{equation}
where only the $u_j$'s located in the disk $D_R$ of radius $R$
centered at the origin contribute to the product and the constant
$C$ is due to the normalization $\int|\psi|^2=1$. For ${\cal
A}>\pi$, we now prove the following result for the atomic density
$\rho_a(\bs r)=|\psi(\bs r)|^2$:
 \begin{equation}
  \label{sigma}
  \rho_a(\bs r) \to p(\bs r)\; \bar \rho_a(r) \quad \mbox{ as }
  R \to \infty\ ,
\end{equation}
where $p(\bs r)$ is periodic over the lattice and vanishes at the
$u_j$'s, and where
\begin{equation}
 \label{sigma2}
 \bar \rho_a(r)=\frac{1}{\pi \sigma^2}e^{-r^2/\s^2}
  \ ,\quad \frac1{\s^2}=1-\frac\pi {\cal A} \ .
  \end{equation}
The function $\bar \rho_a(r)$ is the coarse-grained average of the
atomic density $\rho_a(\bs r)$. This gaussian decay has already
been obtained by Ho in the so-called \emph{averaged vortex
approximation} \cite{H}. However we find useful to prove it here
with a different approach, which we shall generalize to non
uniform lattices (\S~\ref{subsec:distorted}).

To prove this result, we write $\ln(\rho_a(\bs r))=v(\bs r)+w(\bs
r)$, with $w(\bs r)=\ln(\bar \rho_a(\bs r))$ and
 \begin{eqnarray}
v(\bs r) &=& \g'+2 \sum_{u_j\in D_R} \ln|u-u_j| -\frac{2}{\cal
A}\int_{P_R}\ln |u- u'|\;d^2r' \nonumber\\
w(\bs r) &=& -\g-r^2 +\frac{2}{\cal A}\int_{P_R}\ln |u-
u'|\;d^2r'\ ,
 \label{defw}
 \end{eqnarray}
where we set $u'=x'+iy'$, $\g=\ln(\pi \sigma^2)$ and $\g'=\g+\ln
C$. Here $P_R$ denotes the inner surface of the polygon formed by
the union of all elementary cells having their center $u_j$ in the
disk $D_R$. We want to find the limit of $v(\bs r)$ and $w(\bs r)$
as $R$ is large.

We start with the calculation of $v(\bs r)$. The integral entering
in the definition of $v$ can be written
 \begin{equation}
{1\over {\cal A}}\int_{P_R}\ln |u- u'|\;d^2r'= \sum_{u_j\in D_R}
\ffint \ln |u- u_j -u"|\;d^2r"
 \label{integralPR}
 \end{equation}
where the sign $\fint$ stands for the integration over the unit
cell of the lattice divided by the area of the cell ${\cal A}$.
When $R$ tends to infinity, $v(\bs r)$ tends to the series
 \begin{equation}\label{vinfty}
v_\infty(\bs r)=\g_0+ 2 \sum_{u_j} \ffint \ln \frac{|u-u_j|}{|u-
u_j -u"|}\;d^2r"
 \end{equation}
whose convergence can be checked by expanding the function $\ln
|u- u_j -u"|$ up to third order in $u"/(u-u_j)$. This series is a
periodic function over the lattice and we set $p(\bs r)=
\exp(v_\infty(\bs r))$, which is also periodic.

To calculate $w(\bs r)$, we first consider the
auxiliary function $\tilde w(\bs r)= w(\bs r)-w(0)+r^2/\sigma^2$.
Using $\nabla^2 \left[\ln|\bs r-\bs r_0|\right]=2\pi\,\delta(\bs
r-\bs r_0)$, we find that $\tilde w$ is harmonic in
$P_R$, with $\tilde w (0)=0$. Moreover, a small computation leads
to the inequality $\tilde w (\bs r)\geq -\pi r^2/(2{\cal A})$. In
the limit $R\to \infty$, we find that $\tilde w$ converges to
$\tilde w_\infty$, which is a harmonic polynomial with degree less
than 2. Due to the symmetry properties of the unit cell, and the
lower bound by the parabola $-\pi r^2/(2{\cal A})$, $\tilde
w_\infty=0$, hence the result Eq.~(\ref{sigma2}).

To summarize, when the vortex lattice is periodic with a uniform
average spatial density $\bar \rho_v$, the coarse-grain average
$\bar \rho_a$ of the atomic density is the gaussian of
width $\sigma$.  The relation in (Eq.~\ref{sigma2}) can be put
in the form
 \begin{equation}
\nabla^2 [\ln(\bar \rho_a(r))]=-4+4\pi \bar \rho_v\ .
 \label{rhovrhoater}
 \end{equation}
which generalizes to coarse-grained quantities the result given in
Eq.~(\ref{rhovrhoa}). The fast rotation limit corresponds to the
case of a large spatial extent of the atom distribution, i.e. $\s
\to +\infty$  or equivalently ${\cal A}=1/\bar \rho_v \to \pi$.

\subsubsection{The energy associated with a uniform vortex lattice}

Once the behavior of the limiting function $\psi$ is known, we can
determine the reduced energy (\ref{energy3}) in the limit of fast
rotation. This requires the calculation of the integrals $\int
\rho_a(\bs r)\;d^2r$, $\int r^2 \rho_a(\bs r)\;d^2r$ and
$\int\rho_a^2(\bs r)\;d^2r$ in the limit $R\to \infty$. It is
performed using Eq.~(\ref{sigma}) and (\ref{sigma2}), by taking
advantage on the difference in the scales of variations of $\bar
\rho_a(r)$ (scale $\sigma\gg 1$) and $p(\bs r)$ (scale $\sim 1$).
We get \cite{A}
 \begin{equation}
\int \rho_a(\bs r)\;d^2r \simeq \left(\ffint p(\bs r)\;d^2r\right)
\times\left( \int \bar \rho_a(r)\;d^2r \right)
 \label{splitting}
 \end{equation}
so that the normalization of $\rho_a$ entails $\fint p(\bs
r)\;d^2r = 1$. A similar splitting between $p$ and $\bar \rho_a$
occurs for the energy and we find:
\begin{equation}
\epsilon \simeq \int\left(  r^2\,\bar \rho_a(r)
+\frac{b\Lambda}{2}\, \bar \rho_a^2(r)\right)d^2r
=\s^2+\frac{b\Lambda}{4\pi\s^2}
 \label{enerequiv}
\end{equation}
where we have set
 \begin{equation}
 \ffint p^2(\bs r)\; d^2 r= b\ .
 \label{def_b}
 \end{equation}
The reduced energy Eq.~(\ref{enerequiv}) depends on the area
${\cal A}$ of the unit cell through $\s$ and on its shape through
the Abrikosov coefficient $b$.  Let us briefly recall the origin
of this coefficient. Instead of using the exact atomic density
$\rho_a(\bs r)$ to calculate the energy, we work with the
coarse-grain average $\bar \rho_a(r)$, whose spatial variation is
much simpler. To do this substitution, we must renormalize the
interaction coefficient $G$, which becomes $bG$. This is a
consequence of the discreteness of the vortex distribution: since
the wave function $\psi(\bs r)$ must vanish at the vortex
location, the average value of $|\psi|^4$ over the unit cell,
hence the interaction energy, is larger than the result obtained
if $|\psi|$ was quasi-uniform over the cell.

We now look for the choice of $b$ and $\sigma$ which minimizes the
reduced energy Eq.~(\ref{enerequiv}). As known for the case of
superconductors, the lattice minimizing $b$ is the triangular one
\cite{Kleiner64}, for which $b\simeq 1.1596$. The minimization
over $\sigma$ then leads to:
 \beq
 \label{Eopt}
\s_0=(b\Lambda/(4\pi))^{1/4} \hbox{ and
}\epsilon_0=\sqrt{b\Lambda/\pi}.
 \eeq
We recover a scaling similar to
Eqs.~(\ref{ThomasFermi})-(\ref{LLLbound}), inferred for a
distribution varying as an inverted parabola. Note that the size
of the elementary cell ${\cal A}=\pi(1-\sigma_0^{-2})^{-1}$
differs from the rigid body rotation result, ${\cal A}_{\rm
RBR}=\pi/ \Omega$, although the two quantities tend to $\pi$ when
$\Omega$ tends to 1. Actually if we impose ${\cal A}={\cal A}_{\rm
RBR}$ in Eq.~(\ref{enerequiv}), instead of minimizing on $\s$, we
find that $1/\s^2=1-\O$ and we obtain $E_{\rm LLL}\sim 2$, much
larger than the result $E_{\rm LLL}\sim 1$ deduced from
Eq.~(\ref{Eopt}).

The reduced energy $\epsilon_0$ exceeds the lower bound
Eq.~(\ref{LLLbound}) by the factor $\sqrt{b}\times\sqrt{9/8}\sim
1.14$.  The origin of the coefficient $b$ has been explained
above. The coefficient $\sqrt{9/8}=1.06$ is due to the difference
between the gaussian envelope found here (c.f.
Eq.~(\ref{sigma2})), and the optimum function varying as an
inverted parabola Eq.~(\ref{ThomasFermi}). For the parameter
$\Lambda=3000$ used in Fig.~\ref{fig:Photovx}, we find
$\epsilon_0=33.3$ using Eq.~(\ref{Eopt}), which is $\sim 6\%$
larger than the result found numerically  (cf.
Fig.~\ref{fig:Photovx} and \S~\ref{subsec:numerics} below).

\begin{figure}
 \centerline{\includegraphics{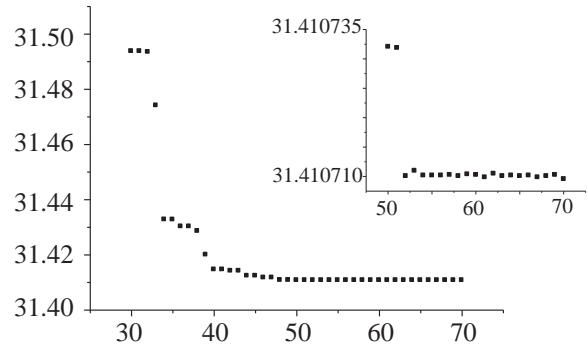}}
 \caption{Minimum reduced energy $\epsilon_n$ as a function of
the number of vortices in the trial wave function
($\Lambda=3000$).} \label{fig:espilonn}
\end{figure}

\subsection{Minimization in the LLL: numerical results}
\label{subsec:numerics}

We now turn to the description of the numerical method that has
been used to obtain the  vortex and atomic patterns shown in
Fig.~\ref{fig:Photovx} and we give some further results of
interest for the following discussion. For a given
$\Lambda=G/(1-\Omega)$ and a given number $n$ of vortices, we
write our trial functions under the form  of Eq.~(\ref{formLLL2}).
We vary the location of the vortices $u_i$ using a conjugate
gradient method to determine the optimal location and the minimum
reduced energy $\epsilon_n$.

The computation of the energy uses the Gauss point method (see
\cite{maday} or \cite{cances} for a use in the case of the Gross
Pitaevski equation): the computation of the integral of a
polynomial times a gaussian is exact as long as the degree of the
polynomial is lower than a certain bound, which depends on the
number of Gauss points. An alternative method used for example in
\cite{KCR} consists in writing the trial functions in the form
Eq.~(\ref{formLLL1}) with $P(u)=\sum_{j=1}^n b_j u^j$, and
performing the minimization by varying the coefficients $b_j$. The
advantage of the method followed here is to give directly the
location of the vortices, while the alternative approach requires
to find the $n$ roots of the polynomial $P(u)$, which may be a
delicate task for large $n$.

\begin{figure}
\includegraphics[width=45mm]{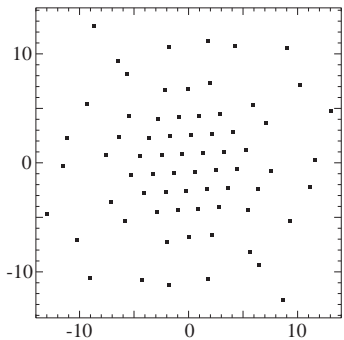}
\caption{Vortex distribution minimizing the reduced energy for
$\Lambda=3000$ and $n=70$ vortices.} \label{photo70}
\end{figure}

For the range of $\Lambda$'s that we have explored (between 300
and 3000), the reduced energy $\epsilon_n$ decreases for
increasing $n$, until it reaches a plateau. For $\Lambda=3000$
(Fig.~\ref{fig:espilonn}), the plateau is reached for $n=52$ and
the reduced energy varies in relative value by $\sim \pm 10^{-8}$
when $n$ increases from 52 to 70. The vortex and atom
distributions for $n=52$ are given in Fig.~\ref{fig:Photovx}. When
$n$ increases the central distribution of vortices remains the
same, as well as the significant part of the atom distribution.
The distribution minimizing the energy for $n=70$ vortices is
shown in Fig.~\ref{photo70}. We note that beyond $n=52$, the
location of the additional vortices strongly depends on the
initial data of the optimization procedure, as extra vortices only
change slightly the energy. In addition to the result of
Fig.~\ref{photo70}, which is the absolute minimum for
$\{\Lambda=3000$, $n=70\}$, we have found a number of
configurations corresponding to local minima where the additional
vortices lie on an outer distorted circle.

\subsection{The distorted lattice}
\label{subsec:distorted}

Inspired by the numerical results such as the ones shown in
Fig.~\ref{fig:Photovx} and Fig.~\ref{photo70}, we generalize the
approach developed for the regular vortex lattice to the case of a
distorted lattice. We make the hypothesis that the locations $u_j$
of the vortices are deduced from a regular hexagonal lattice
$u_j^{\rm reg}$ by
 \begin{equation}
 u_j=T(u_j^{\rm reg}) =\la(|u_j^{\rm reg}|)\, u_j^{\rm reg}
 \label{distorted}
 \end{equation}
where $\lambda(r)$ is a positive function varying smoothly over a
distance of order unity. We assume that the unit cell of the
initial regular lattice has the area ${\cal A}=\pi$, corresponding
to a flat density profile in Eq.~(\ref{sigma2}). If $\lambda$
tends to infinity for a finite value $r_h$, the number of vortices
in the distorted lattice is finite and equal to $\sim r_h^2$,
since all vortices located after the `horizon' $r_h$ in the
regular lattice are rejected to infinity. Otherwise, if $\lambda$
is finite for all $r$, the number of vortices in the distorted
lattice is infinite.

\begin{figure}
\centerline{\includegraphics[width=85mm]{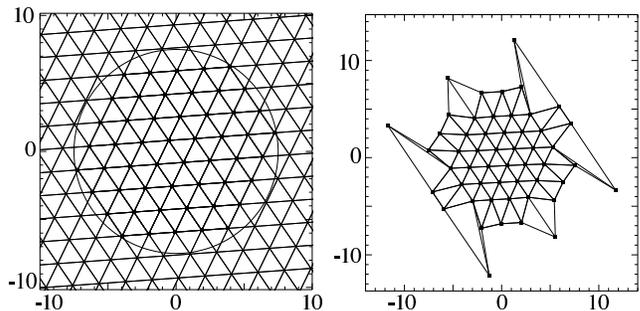}}
 \caption{Regular lattice with ${\cal A}=\pi$ and distorted lattice
 minimizing the energy for $\Lambda=3000$ and $n=52$ vortices.}
\label{fig:res}
\end{figure}

The distortion is illustrated in Fig.~\ref{fig:res} for the
particular case of $\Lambda=3000$. We have plotted at the same
scale the regular lattice with ${\cal A}=\pi$ and the
configuration of vortices minimizing the energy. For $n=52$
vortices, only the lattice sites of the regular lattice whose
distance to the origin is below $r_h=7.4$ remain in the distorted
lattice. Around the center of the disk of radius $r_h$, the
function $\lambda(r)$ is close to 1, whereas it becomes very large
when $r$ approaches the `horizon' $r_h$.

As for the case of the regular lattice, we introduce the
coarse-grained averages $\bar \rho_a$ and $\bar \rho_v$ of the
atom and vortex densities. The function $\bar \rho_v$ is now space
dependent and is simply the inverse of the area of a distorted
cell in the vicinity of $\bs r$:
 \begin{equation}
\bar \rho_v(\bs r)=\left( \pi \la(r')(\la(r')+r'\la'(r'))
\right)^{-1}\ ,
 \label{rhovbar0}
 \end{equation}
where $\bs r= \la(r') \bs {r'}$. We recall that the expected
length scale in the limit of fast rotation is
$R_0=(2\Lambda/\pi)^{1/4}\gg 1$ (see \S~\ref{subsec:scaling}) and
we consider a class of distortions $\lambda(r)$ such that
\begin{equation}
 \label{la2}
 \la^2(r)=1+\frac{f(r^2/R_0^2)}{R_0^2}+O\left (
 \frac{1}{R_0^4}\right )\ ,
\end{equation}
where $f(\xi^2)$ is a continuous function, which diverges at
$\xi_h^2=r_h^2/R_0^2$. We also assume that the integral
$F(s)=\int_0^s f(s')\;ds'$ diverges at $s=\xi_h^2$.  We shall
check in the end that the distortion minimizing the energy belongs
to the class of functions defined in Eq.~(\ref{la2}).

In the limit $\Lambda \gg 1$, we show in the appendix the
following properties for the vortex lattices obtained through a
distortion obeying (Eq.~\ref{la2}):
\begin{enumerate}
\item
The atom density $\rho_a(\bs r)$ can be written as
 \begin{equation}
\ln\left( \rho_a(\bs r)\right)=v(\bs r)+w(r)
 \end{equation}
where $v$ is related to the function $v_\infty(\bs r)$ introduced
for a regular lattice in Eq.~(\ref{vinfty}):
 \begin{equation}
v(\bs r)=v_\infty(\bs r') \quad \mbox{with}\quad \bs
r=\lambda(r')\bs r'\ .
 \end{equation}
$w(r)$ is a smooth radial function and we set $\bar
\rho_a(r)=\exp(w(r))$, where $\bar \rho_a$ is normalized to unity.

\item The coarse-grain average $\bar \rho_a$ at a point $\bs
r=R_0\bs \xi$ and the integral $F$ of the distortion function $f$
are related by the relation:
 \begin{equation}
\bar \rho_a (R_0 \xi) \propto \exp(-F(\xi^2)) \quad \mbox{if}\quad
\xi< \xi_h\ , \label{Frhoa}
 \end{equation}
and is zero elsewhere.  Note that $\bar \rho_a(r)$ is continuous
at $r_h=R_0\xi_h$ since we have assumed that $F(s)$ tends to
$+\infty$ when $s\to \xi_h^2$.

\item As for the regular lattice case, we use the difference in
the scales of variations of the two functions $v$ and $w$ to
obtain
 \begin{equation}
\epsilon \simeq \int \left(  r^2\,\bar \rho_a( r)
+\frac{b\Lambda}{2}\, \bar\rho_a^2(r)\right)d^2r \ .
 \label{E_distorted}
 \end{equation}
where $b=1.1596$ as for a regular lattice.
\end{enumerate}

The differences with respect to the initial minimization problem
of Eq.~(\ref{energy3}) are the renormalization of the coefficient
$G\to bG$ discussed in \S~\ref{subsec:regular}, and the fact that
$\bar \rho_a$ is a smooth,  non-negative radial function, instead
of being the square of an LLL wavefunction.

\subsection{The ``Thomas-Fermi" distribution in the LLL}
\label{subsec:TFLLL}

We now address the minimization of the energy functional in
Eq.~(\ref{E_distorted}). The minimizing function is the inverted
parabola $\bar \rho_a(r) \propto R_1^2-r^2$ for
$r<R_1=(2b\Lambda/\pi)^{1/4}$, and $\bar \rho_a=0$ for $r>R_1$.
The associated energy is
 \begin{equation}
\epsilon \simeq  \frac{2\sqrt{2}}{3\sqrt{\pi}} \sqrt{b\Lambda}.
 \label{energy4}
  \end{equation}
Using Eq.~(\ref{Frhoa}) we deduce  the distortion function $f(s)$
and its primitive $F(s)$:
 \begin{equation}
 f(s) \simeq \frac{1}{\sqrt{b}-s}\qquad
 F(s)=-\ln\left( 1-\frac{s}{\sqrt b} \right)
 \label{distortionf}
 \end{equation}
As initially assumed, the functions $f(s)$ and $F(s)$ tend to
$+\infty$ at the horizon $\sqrt{b}$, hence $r_h=b^{1/4}R_0=R_1$.
This means that at leading order in $\Lambda$, the Thomas Fermi
radius and the horizon are equal. The function $T$ transforming
the initial regular lattice $u_j^{\rm reg}$ into the distorted
lattice $u_j$ is thus:
 \begin{equation}
\bs r= T(\bs r')=\bs r'+\frac{\bs r'}{R_1^2-r'^2}\ .
 \label{transform}
 \end{equation}
 Once $f(s)$ is known, one
can evaluate the vortex density using Eq.~(\ref{rhovbar0}):
 \begin{equation}
 \pi\bar\rho_v(r)=\left(1+\frac{R_1^2}{(R_1^2-r'^2)^2}\right)^{-1} \ .
 \label{rhovbar}
 \end{equation}
where $\bs r$ and $\bs r'$ are related by Eq.~(\ref{transform}).
In particular the vortex density at $r=0$ is $\sim
(1-R_1^{-2})/\pi$, which is close, but not equal, to the
prediction $\Omega/\pi$ for a rigid body rotation.

Our distortion function $f$ is to be related to that of \cite{AC},
though it is derived using very different techniques. The
asymptotic result Eq.~(\ref{energy4}) has also been obtained
recently by Watanabe, Baym and Pethick \cite{WBP} who  assumed
that Eq.~(\ref{rhovrhoater}) can be generalized to the case where
$\bar \rho_v$ is spatially dependent:
 \begin{equation}
\nabla^2 [\ln(\bar \rho_a(r))]=-4+4\pi \bar \rho_v (r)\ .
\label{rhovrhoaquat}
 \end{equation}
 By differentiating Eq.~(\ref{Frhoa}), a similar relation can be
proved within our approach with $\bar \rho_v (r)$ replaced by
$\bar \rho_v (T(r))$. The two relations are equivalent at points
not too close to the Thomas-Fermi radius (i.e. $R_1-r \gtrsim 1$).
A result related to Eq.~(\ref{rhovrhoaquat}) has also been shown
in a different context by Sheehy and Radzihovsky \cite{S}. They
consider the case of a condensate which is not in very fast
rotation (i.e. outside of the LLL regime) but still with several
vortices. Interestingly, the procedure used in \cite{S} to derive
the relation between $\bar \rho_v$ and $\bar \rho_a$ is based on
the minimization of the energy functional, including atom
interactions. On the contrary, the result in Eq.~(\ref{Frhoa}) or
Eq.~(\ref{rhovrhoaquat}) is a consequence of the structure of an
LLL wave function and it is at first sight independent of atomic
interactions. However one must keep in mind that the knowledge of
the strength of atom interactions is essential to check the
relevance of LLL wave functions for the problem (see
Eq.~(\ref{LLLfrontier})).  The relation reached in \cite{S} has
the same structure as Eq.~(\ref{rhovrhoaquat}), but with a
dimensionless coefficient involving the healing length and $\bar
\rho_v$ inside the $\nabla^2 \ln(\rho_a)$ term.  Close to the
Thomas-Fermi radius, $\bar \rho_v$ varies rapidly and the approach
of \cite{S} leads to a different relation from
Eq.~(\ref{rhovrhoaquat}), since the derivatives of $\bar\rho_v$
have a significant contribution in this region.

\begin{figure}
\centerline{\includegraphics[width=60mm]{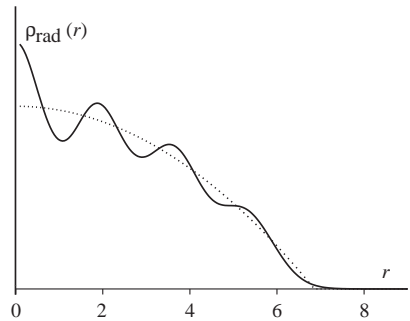}}
 \caption{Radial
density distribution $\rho_{\rm rad}(r)$ for $\Lambda=3000$. The
unit along the vertical direction is arbitrary. The dotted line is
a fit using the inverted parabola with the radius
$R_1=(2b\Lambda/\pi)^{1/4}$, with an adjustable amplitude.}
\label{fig:distribrad}
\end{figure}

Our analytical predictions can be compared with our numerical
results obtained in the particular case $\Lambda=3000$ (i.e.
$R_1=6.86$), for which we plotted Fig. \ref{fig:Photovx}. The
prediction of Eq.~(\ref{energy4}) yields $\epsilon=31.374$, only
$0.12\%$ below the value determined numerically. We can also
compare our trial density with the numerical result. We give in
Fig.~\ref{fig:distribrad} the prediction of the inverted parabola
together with the radial density distribution determined
numerically:
 \begin{equation}
\rho_{\rm rad}(r)=\frac{1}{2\pi} \int_0^{2\pi} |\psi(\bs r)|^2\;
d\theta
 \end{equation}
where $\theta$ is the polar angle in the $xy$ plane. Apart from
oscillations due to the discreteness of vortices, the two
distributions are remarkably close to each other. A similar
conclusion was reached recently by Cooper, Komineas, and Read
\cite{KCR}. They also performed a numerical minimization of the
energy of Eq.~(\ref{energy1}) in the LLL limit, and found an atom
density profile in good agreement with the inverted parabola
distribution predicted in \cite{WBP}.

 From the above analytical results, we expect that the minimizing
configuration will involve $n\sim r_h^2\sim 48$ vortices. The
number of vortices for which the minimum energy plateau is reached
numerically is 52, which is very close to $r_h^2$.  As for the
location of vortices, our analysis indicates that the vortices in
the distorted lattice are images through
Eqs.~(\ref{distorted})-(\ref{transform}) of points of the regular
hexagonal lattice such that $|u_j^{\rm reg}|<r_h=R_1$.   Note that
the optimal vortex configuration involves some vortices outside
the disk of radius $R_1$. They correspond to regular lattice sites
$|u_j^{\rm reg}|$ close to the horizon $r_h$. Indeed, for these
points $\la$ gets large and the image point is sent beyond the
Thomas Fermi radius. Thus, though the distorted lattice provides
an inverted parabola which vanishes at $R_1$, the location of the
vortices extends beyond $R_1$. The numerical analysis leads to
results which nicely confirm our analytical predictions. In
addition it allows to explore the role of the vortices lying
outside the Thomas-Fermi distribution. For example one can remove
the contribution $(u-u_j)$ of these vortices in the expression
Eq.~(\ref{formLLL2}) of the LLL wave function, while keeping
unchanged the contribution of the vortices inside the Thomas-Fermi
radius. This results in a significant modification of $\rho_a(\bs
r)$ which then vanishes around $\sim 7.3$, instead of $\sim 6.8$.
Therefore these outer vortices play an important role in the
equilibrium shape of the condensate, even though they cannot be
found when one simply plots the atomic spatial density.

A closer look at Fig.~\ref{fig:distribrad} indicates that $\bar
\rho_a$ is matched to zero more smoothly than an inverted
parabola. An expansion of the energy of the distorted lattice
(Eq.~(\ref{E_distorted})) to the next order in $\Lambda$ should
lead to a minimizing function $\bar \rho_a$ with a smoother decay
to zero around $R_1$. In particular a natural way to match the
inverted parabola with the asymptotic decay $r^{2n}e^{-r^2}$ of
any LLL function with $n$ vortices, could be obtained through a
Painlev{\'e}-type equation (as at the border of a non-rotating
BEC).

\paragraph*{Remark: Comparison with the ``centrifugal force approximation".}
Under some conditions, it is possible to write $\psi$ as the
product of a rapidly varying function $\eta(\bs r)$ and a slowly
varying envelope $\bar \psi(\bs r)$ \cite{BP}. This is reminiscent
of the splitting of $\ln (\rho_a)$ in terms of $v$ and $w$,
although it leads to a different conclusion. One obtains for the
envelope an equation similar to Eq.~(\ref{GP1}), where only the
centrifugal potential remains \cite{BP}:
 \begin{equation}
-\frac{1}{2}\nabla^2 \bar \psi(\bs r) + (1-\Omega^2) \frac{r^2}{2}
\bar \psi(\bs r) +G|\bar \psi(\bs r)|^2 \bar \psi(\bs r)=\bar \mu
\,\bar \psi(\bs r)
 \label{centrifug}
 \end{equation}
where $\bar \mu=\mu -\Omega$. We call this approach the
``centrifugal force" approximation \cite{renorm} and we compare
its predictions with those derived from the LLL approximation.

When the approximation leading to Eq.~(\ref{centrifug}) is valid,
one is left with the problem of a 2D gas at rest in a harmonic
potential with the spring constant $1-\Omega^2$. The solution of
this equation depends on the strength of the interaction parameter
$G$. If $G\gg 1$, the kinetic energy term can be neglected
(Thomas-Fermi approximation) and one gets $ |\bar \psi(\bs r)|^2
\propto 1-{r^2}/{R_{\rm cfa}^2}$ inside the disk of radius $R_{\rm
cfa}= \left({4G}/[{\pi(1-\Omega^2)]}\right)^{1/4}$ and $\psi(\bs
r)=0$ outside. Note that $R_{\rm cfa}$ coincides with our
Thomas-Fermi radius $R_1$ for $\Omega\simeq 1$. If $G \ll 1$, the
interaction term can be neglected and the solution is the ground
state of the harmonic oscillator, i.e. the gaussian of width
$(1-\Omega^2)^{-1/4}$.

In the LLL, we have seen that the distinction between the two
regimes $G\gg1$ and $G\ll 1$ is not relevant. The only important
parameter is $\Lambda=G/(1-\Omega)$. When $\Lambda \gg 1$ the
envelope of the atom density profile is close to an inverted
parabola, irrespective of the value of $G$. Therefore, there
exists a clear discrepancy between the predictions of the LLL
treatment and those of the centrifugal force approximation when
$1-\Omega \ll G \ll 1$. For these parameters the LLL approximation
is valid since $G(1-\Omega) \ll 1$ (see Eq.~(\ref{LLLfrontier})).
The extent of the wave function minimizing $\epsilon[\psi]$ is
thus $R_1\sim (G/(1-\Omega))^{1/4}$, while the reasoning based on
Eq.~(\ref{centrifug}) would lead to a gaussian envelope with a
larger size $(1-\Omega)^{-1/4}$, independent of $G$.


\section{Extension to other confining potentials}
\label{sec:extension}

The ideas that we have developed for a harmonic confinement can be
generalized to a larger class of trapping potential $V(\bs r)$.
For simplicity we assume here that $V$ is cylindrical symmetric,
with a minimum at $r=0$. We define $\omega$ as
$m\omega^2=\partial^2 V/\partial r^2|_0$ and we set
 \begin{equation}
V(r)=\frac{1}{2}m\omega^2 r^2 + W(r)\ .
 \end{equation}
As above we choose $\omega$ and $\sqrt{\hbar/(m\omega)}$ as the
units for frequency and length, respectively.

We are still interested here in a region where $\Omega\sim 1$. To
minimize the Gross-Pitaevskii energy functional, we use again wave
functions in the LLL so that the energy per particle to be
minimized is
 \begin{equation}
E_{\rm LLL}=\Omega+\int \left[ \left((1-\Omega)r^2 +W(r) \right)
\rho_a +\frac{G}{2}\rho_a^2\right]\,d^2r
 \label{energy5}
 \end{equation}
As explained in section II.C, the LLL approximation is valid if
the minimum for $E_{\rm LLL}-\Omega$ is small compared to the
distance $2=2\hbar \omega$ between the LLL and the first excited
Landau level.

We have seen that varying the locations $u_i$ of the vortices,
hence the average vortex surface density $\bar \rho_v$, allows to
generate a large class of coarse-grain averaged atom densities
$\bar \rho_a$. Provided $W(r)$ is well behaved, we can generalize
the treatment presented for the purely quadratic case. The energy
$E_{\rm LLL}$ can still be expressed in terms of $\bar \rho_a$
instead of $\rho_a$ with an expression similar to
Eq.~(\ref{energy5}), and the interaction parameter $G$ replaced by
$bG$. The Thomas-Fermi distribution minimizing $E_{\rm LLL}$ is
 \begin{equation}
 \label{tra}
\rho_a^{\rm TF}(r) = \max\left( \frac{\mu-(1-\Omega)r^2-
W(r)}{bG}, 0\right)
 \end{equation}
where $\mu$ is the chemical potential determined such that $\int
\bar \rho_a=1$. Once $\bar \rho_a$ has been determined over the
whole space, the energy $E_{\rm LLL}$ can then be calculated and
the validity of the various approximations can be checked: (i)
$|E_{\rm LLL}-\Omega|\ll 1$ and (ii) the extension of the domain
where $\bar \rho_a$ differs from zero is large compared to 1, so
that it is legitimate to introduce a coarse-grain average of
$\rho_a$ over several vortex cells, and there is a large parameter
playing the role of $R_0$.

As an example, we investigate the case of a combined quartic and
harmonic potential: $W(r)=kr^4/4$, which has been studied recently
both theoretically, numerically
\cite{Fetter01,Kasamatsu02,L,Kavoulakis03,Aftalion03,JKL,Fetter04}
and experimentally \cite{Bretin04}. A nice feature of this
potential is that it allows to explore the region $\Omega\geq 1$,
since the centrifugal force, $-\Omega^2 r$, can always be
compensated by the trapping force, varying as $-(r+kr^3)$. We
define $\Delta_0=(3k^2bG/(8\pi))^{2/3}$ and
$\Delta=(1-\Omega)^2+k\mu$. Two cases can occur. (i) If
$\Omega<\Omega_c=1+\sqrt\Delta_0$, then $\rho_a^{\rm TF}$ is non
zero in a disc of radius $R_+^2=2(\Omega-1+\sqrt\Delta)/k$,
$\Omega$ and $\Delta$ being linked by
\begin{equation}
 2\Delta^{3/2} + 3\Delta (\Omega-1)- (\Omega-1)^3=4\Delta_0^{3/2} \ .
 \end{equation}
(ii) If $\Omega>\Omega_c$, then $\rho_a^{\rm TF}$ is non-zero on
an annulus of radii $R_\pm^2=2(\Omega-1\pm\sqrt\Delta) /k$, and
$\Delta=\Delta_0$.

The Thomas-Fermi distribution  given in Eq.~(\ref{tra}) allows to
calculate the minimum energy per particle. Since the general
calculation is quite involved, we simply give here the result for
$\Omega=\Omega_c$:
 \begin{equation}
\Omega=\Omega_c:\quad E_{\rm LLL}-\Omega=\alpha\,k^{1/3}\,G^{2/3}
 \end{equation}
where $\alpha\simeq -0.1$. More generally, when $|1-\Omega|$ is at
most of the order of $k^{2/3}G^{1/3}$, then $E_{\rm LLL}-\Omega$
is of the order of $k^{1/3}\,G^{2/3}$. The restriction to the LLL
wave functions and the use of the `Thomas-Fermi' approximation
(Eq.~(\ref{tra}) are valid if two conditions are fulfilled: (i)
$E_{\rm LLL}-\Omega\ll 1$, hence $k G^2\ll 1$, (ii) the extension
$R_+\sim (G/k)^{1/6}$ of $\bar \rho_a$ is large compared to 1, so
that the coarse-grain average of $ \rho_a$ is meaningful. This
requires $k\ll 1$ and $k \ll G \ll 1/\sqrt{k}$. When these
conditions are satisfied, $\Omega_c -1 \sim k^{2/3}G^{1/3} \ll 1$,
and the study of the regime $\Omega \geq \Omega_c$ can be
performed within the LLL. In addition one can check that for
$\Omega_c-1 < \Omega-1 \ll k^{1/3}G^{2/3}$, the width $R_+-R_-$ of
the annulus is large compared to 1 (both $R_+$ and $R_-$ are of
order $(G/k)^{1/6}$), so that the use of the coarse-grain averages
of $\rho_a$ and $\rho_v$ is justified. A similar analysis to what
we have performed above yields an almost uniform vortex lattice in
the annulus, with a distortion near the inner and outer
boundaries.

The LLL approximation has been used by Jackson, Kavoulakis and
Lundh to study the phase diagram of the vortices in a
quadratic+quartic phase \cite{JKL}. They were mostly interested in
the stability of giant vortices, hence they restricted their
analysis to particular LLL states, where $F(u)$ only contains two
or three terms $b_j u^j$. However one could in principle use the
same approach as the numerical treatment developed here, and
derive the detailed vortex pattern for various choices of $G$,
$\Omega$ and $k$. It would be interesting to see whether there
exists a domain of parameters where the polynomial $F(u)$ has a
multiple root in $u=0$. This would correspond to the giant vortex
which has been predicted by other approaches
\cite{L,Fetter04,Fischer03}. Another limit where $R_+-R_- \leq 1$
has recently been studied in \cite{Fetter04}.


\section{Conclusion}
\label{sec:conclusion}

In this paper, we have studied analytically and numerically the
vortex distribution and atomic density for the ground state of a
rotating condensate trapped in a harmonic potential, when the
rotation and trapping frequencies are close to each other.
Restricting our analysis to quantum states in the lowest Landau
level, we have shown that the atomic density varies as an inverted
parabola over a central region. The vortices form an almost
regular triangular lattice in this region, but the area of the
cell differs from the prediction for solid body rotation. In the
outer region, the lattice is strongly distorted. We have
determined the optimal distortion, and related it to the decay of
the wave function close to the Thomas-Fermi radius.

Our results agree with those of a recent numerical study
\cite{KCR}. Another analytical approach to this problem has
recently been given in \cite{WBP}. It leads to the same value as
ours for the energy of the ground state, whereas our treatment
provides more detailed information on the vortex pattern at the
edge of the condensate. Our predictions for the equilibrium shape
of the atomic density and for the vortex distribution should be
experimentally testable. In \cite{Boulder04} the regime of fast
rotation in the LLL has already been achieved and it was indeed
found that the atom density profile varies as an inverted
parabola, and not as a Gaussian as one would expect for an
infinite regular lattice \cite{H}. In \cite{Boulder04bis}, a
detailed experimental analysis of the vortex spacing as a function
of the distance to the center of the trap has been made and it
showed a clear distortion of the pattern on the edges of the
condensate. This study was not performed in conditions such that
our LLL approximation is valid, and the relevant theoretical model
is rather the one developed in \cite{S}. However it should be
possible to perform a similar experimental analysis for faster
rotation rates, and test in particular the validity of our
prediction concerning the distortion factor $\lambda(r)$ (see
Eqs.~(\ref{distorted})-(\ref{la2})-(\ref{distortionf})).

Finally, we have addressed the case of other trapping potentials,
such as a superposition of a quadratic and a quartic potential,
which have also been addressed experimentally \cite{Bretin04}. For
even faster rotations, when the number of vortices approaches the
number of atoms, the ground state is strongly correlated. We did
not touch this point here, but our work should be relevant for
studying the apparition of this correlated regime from a
destabilization of the mean field results by quantum fluctuations.

\acknowledgements

J.D. is indebted to Yvan Castin and Vincent Bretin for several
insightful discussions. A.A. and X.B. are very grateful to Fran\c
coit Murat for explaining details on homogeneisation techniques
and to Eric Canc{\`e}s for pointing out the method of Gauss points
in the computations. This work is partially supported by the fund
of the French ministry for research, ACI ``Nouvelles interfaces
des math{\'e}matiques'', CNRS, Coll{\`e}ge de France, R{\'e}gion
Ile de France, and DRED.


\section{Appendix}
\label{sec:appendix}

\begin{figure}
\centerline{\includegraphics[width=65mm]{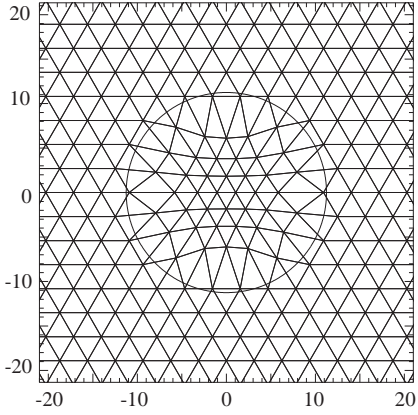}}
\caption{Example of a distorted lattice generated by the
transformation Eq.~(\ref{tl}). The radius of the circle is
$\lambda_\alpha \, \alpha R_0$. In the regular part outside the
circle, the cell area is ${\cal A}_\alpha$.}
 \label{fig:distort}
\end{figure}

The aim of this appendix is to prove the properties used in
\S~\ref{subsec:distorted}. A detailed proof will be given in
\cite{AB}. We consider a distorted lattice in an inner region and
keep a regular lattice in the outer region in such a way that the
distortion is continuous (see Fig.~\ref{fig:distort}). We label by
$\bs j$ a regular hexagonal lattice with a unit cell area ${\cal
A}=\pi$ and we define the transformed lattice by
 \begin{equation}
 \label{tl}
u_j = \left\{
\begin{array}{l}
\la(|\bs j|)\,{\bs j} \quad    \mbox{for} \quad |\bs j|<\a R_0 \\
\la_\alpha\,{\bs j} \qquad\ \hbox{for} \quad |\bs j|\geq \a R_0
 \end{array}
  \right.
 \end{equation}
where the radius $R_0$ is given in Eq.~(\ref{ThomasFermi}), the
distortion function $\la(r)$ satisfies (\ref{la2}), and $\alpha
R_0$ is smaller than the horizon $r_h$ where $\lambda(r)$
diverges. We have set $\lambda_\alpha=\lambda(\alpha R_0)$ and the
area of the unit cell of the outer lattice is ${\cal
A}_\alpha=\pi\la^2_\a$. When $\alpha R_0$ tends to $r_h$, ${\cal
A}_\alpha$ tends to infinity and the vortex lattice of
Fig.~\ref{fig:distort} is similar to the one in the right of
Fig.~\ref{fig:res}.

In the following, we shall (i) define $\bar \rho_a$ and compute
its limit when $R_0$ increases (i.e. $\O$ tends to 1) for a fixed
$\alpha$, (ii) let $\a$ get close to the horizon $\xi_h=r_h/R_0$.
We need to take the limits in this order, because we will use that
$\la_\alpha$ is close to 1, which is only true if $\a$ is fixed
less than $\xi_h$ and $R_0$ is large.

Firstly we consider only the points $\bs j$ in a disc $D_{R'}$
 \begin{equation}
\psi(\bs r)=C e^{-r^2/2} \prod_{|\bs j|<R'} (u-u_j)\ .
 \end{equation}
 $Q_\alpha$
denotes the unit cell of the lattice of area ${\cal A}_\alpha$ and
$P_{\alpha, R'}$ is the polygon formed by the union of all
elementary cells of area ${\cal A}_\alpha$ and center
$\lambda_\alpha \bs j$, with $|\bs j|< R'$. We write
$\ln(\rho_a(\bs r))=v_{R'}(\bs r)+w_{R'}(\bs r)$ with,
 \begin{eqnarray}
v_{R'}(\bs r) &=& 2 \sum_{\a R_0<|\bs j|< R'}\Bigl ( \ln|\bs
r-\lambda_\alpha \bs j|
 \nonumber
 \\
&-&\frac{1}{\cal A_\alpha}\int_{Q_\alpha}\ln |\bs r- \bs
r'-\lambda_\alpha
\bs j|\;d^2r'\Bigr ) \nonumber\\
&+& 2 \sum_{|\bs j|< \a R_0}\Bigl ( \ln|\bs r-\la(\bs j) \bs j|\nonumber\\
&-&\frac{1}{\cal A_\alpha}\int_{Q_\alpha}\ln |\bs r- \bs r'-\la(
\bs j )\bs j|\;d^2r'\Bigr )
 \nonumber
 \end{eqnarray}
and $w_{R'}(\bs r)=w_{1R'}(\bs r)+w_2(\bs r)$ with
 \begin{eqnarray}
w_{1R'}(\bs r)&=&-r^2+\frac{2}{\cal A_\alpha}\int_{P_{\alpha,
R'}}\ln |\bs r- \bs r'|\;d^2r'
 \nonumber
 \\
w_2(\bs r) &=& \sum_{|\bs j|< \a R_0}  \frac{2}{\cal A_\alpha}
\int_{Q_\alpha}\ln \frac{|\bs r- \bs r'-\la( \bs j )\bs j|}{|\bs
r- \bs r'-\lambda_\alpha\bs j|}\;d^2r' \nonumber \label{defw2}
 \end{eqnarray}
We have just added and subtracted  terms at this stage.

Now we let $R'$ tend to infinity and find the limit for an
infinite number of vortices. This step is very similar to the case
of the regular lattice since the lattice distortion only affects a
finite number of sites. We find that
 \begin{equation}
 w_{1R'} (\bs r)-w_{1R'} (0)\ \to\  w_1(\bs r)=-r^2/\s^2
 \end{equation}
with $\sigma^{-2}=1-\pi/{\cal A}_\alpha$. $v_{R'}$ tends to a
convergent series $v$ (which is not a periodic function, contrary
to the regular lattice case).

The next step is to let $R_0$ be large, keeping $\a$ fixed, so
that $\la_\a$ is close to 1 for the class of distortion functions
considered in Eq.~(\ref{la2}). We find
 \begin{equation}
 v(\bs r) \simeq v_\infty(\bs r' )
 \end{equation}
where $v_\infty$ is given by Eq.~(\ref{vinfty}), and $\bs r,\bs
r'$ are related by
 \begin{equation}
\bs r= \left\{ \begin{array}{l}
 \lambda(r')\,\bs r' \qquad \mbox{for}
 \quad r /\lambda_\alpha \leq \alpha R_0\\
 \lambda_\alpha \,\bs r' \qquad \quad
 \mbox{for}
 \quad r /\lambda_\alpha>\alpha R_0\ , \\
 \end{array}
 \right.
 \label{rrprime}
 \end{equation}

We denote $w=w_1+w_2=\ln (\bar \rho_a)$. We estimate $w_2(r)$,
using an expansion of the logarithm and the fact that $\la(\bs j)
\sim \lambda_\alpha\sim 1$:
\begin{equation}
w_2(R_0\bs \xi) \simeq \frac{1}{{\pi}} \int_{\xi'<\alpha}
\left(f(\a^2)-f(\xi'^2)\right) \frac{\bs{\xi'}\cdot (\bs \xi-\bs
\xi')}{|\bs \xi-\bs \xi'|^2}\,d^2\xi'\ .
 \label{defw3}
 \end{equation}
where relevant $\xi$'s are of order unity. Using an integration by
part and a primitive $F$ of $f$, we get ($\theta(x)$ is the
Heaviside function):
\begin{equation}
w_2(R_0\bs \xi) \simeq \left[F(\a^2)-F(\xi^2)+(\xi^2-\a^2)f(\a^2)
\right]\;\theta(\alpha-\xi)
 \label{defw4}
 \end{equation}
Since we have  $\sigma^{-2}\simeq f(\a^2)/R_0^2$, then $w_1(R_0
\bs \xi)=-\xi^2 f(\a^2)$. Putting everything together, we obtain
up to an additive constant for normalization,
 \begin{equation}
\ln(\bar \rho_a(R_0\bs \xi)) \simeq \left\{
 \begin{array}{l}
-F(\xi^2) \qquad  \quad\ \  \mbox{for} \quad \xi<\alpha \\
-\xi^2 f(\a^2)+\mu\quad \mbox{for}\quad \xi>\alpha
 \end{array}
 \right.
 \label{relwF}
 \end{equation}
with $\mu=\a^2f(\a^2)-F(\a^2)$.

Finally, using that $\la_\a \simeq 1$, we can apply the separation
of integrals \cite{A} and find for example that
 \begin{eqnarray}
 \label{L2ra2}
 \int \rho_a d^2r
&\propto& \left(\ffint e^{v_\infty(\bs r')}\;d^2r'\right)
\times \nonumber\\
&&\left( \int_{\xi < \alpha} e^{-F(\xi^2)}\;d^2\xi + e^{\mu}
\int_{\xi>\alpha} e^{-\xi^2f(\a^2)}\;d^2\xi\right)\nonumber
 \end{eqnarray}
The last integral in the second line is equal to $\pi
e^{-F(\a^2)}/f(\a^2)$. At this stage, $\a$ is still a free
parameter. If the distortion function $\lambda(r)$ has a horizon
at $r=\xi_h R_0$, we let $\a$ tend to $\xi_h$, otherwise to
$\infty$. The last integral tends to zero given the hypothesis
that $f$ and $F$ tend to $+\infty$ at $\xi_h$. The same procedure
is valid for all terms entering into the energy functional, which
justifies the use of Eq.~(\ref{E_distorted}).



\begin{thebibliography}{99}
\bibitem{renorm}
Strictly speaking, as shown in \cite{BP}, the coefficient $G$ is
renormalized in this procedure. Here we omit this change for our
qualitative discussion.
\end{thebibliography}
\end{document}